\documentstyle[preprint,prd,aps,amsfonts,epsf]{revtex} % 12 pt
\tightenlines

\newcommand{\sigmaBr}{
	0.132 ^{+0.041}_{-0.037} \, ({\rm stat.}) \, 
		\pm 0.031 \, ({\rm syst.}) \,
	  ^{+0.032}_{-0.020} \, ({\rm lifetime})
}
\newcommand{\ctauBc}{
	137^{+53}_{-49}\,({\rm stat.}) 	\pm 9\,({\rm syst}) \: \mu m
}

\newcommand{\tauBc}{
	0.46^{+0.18}_{-0.16}\, ({\rm stat.}) \, 
	\pm 0.03 \, ({\rm syst.}) \: {\rm ps }
}

\newcommand{\Bc}{$B_c$}
\newcommand{\Bcp}{$B_c^+$}

\newcommand{\Jpsi}{$J/\psi$}

\newcommand{\Jpsimu}{$J/\psi \, \mu $}
\newcommand{\Jpsie}{$J/\psi  \, e   $}
\newcommand{\Jpsil}{$J/\psi  \, \ell$}

\newcommand{\Jpsilnu}{$J/\psi  \, \ell \, \nu$}

\newcommand{\Jpsimumu}{$J/\psi \rightarrow  \mu^+ \mu^-$}

\newcommand{\BJpsienu}{$B_c  \rightarrow J/\psi \, e    \, \nu$}
\newcommand{\BJpsilnu}{$B_c  \rightarrow J/\psi \, \ell \, \nu$}
\newcommand{\BpJpsimunu}{$B_c^+ \rightarrow J/\psi \, \mu^+  \, \nu$}
\newcommand{\BpJpsienu}{$B_c^+  \rightarrow J/\psi \, e^+    \, \nu$}
\newcommand{\BpJpsilnu}{$B_c^+  \rightarrow J/\psi \, \ell^+ \, \nu$}

\newcommand{\BJpsiK}{$B^+  \rightarrow J/\psi \, K^+$}

\newcommand{\cc}{$c$}

\newcommand{\bbar}{$\overline{b}$}

\newcommand{\PbarP}{$p\overline{p}$}
\newcommand{\pbarp}{$p\overline{p}$}

\newcommand{\Pt}{$p_T$}

\newcommand{\ipb}{pb$^{-1}$}

\newcommand{\um}{$\mu$m}

\newcommand{\ctau}{$c\tau$}

\newcommand{\ctstar}{$ct^{\ast}$}

\newcommand{\sigBRBcl}{$\sigma\cdot BR(B_c^+\rightarrow J/\psi\,\ell^+ \nu)$}

\begin{document}
%\draft
\preprint{\begin{minipage}[t]{3in} 
		\flushright
		FERMILAB-PUB-98/157-E
%	\\	CDF/PUB/BOTTOM/CDFR/4496 \vspace*{-10pt}
%	\\	aps1998mmmdd\_yyy \vspace*{-10pt} 
%	\\	PRL Version 4.1 \vspace*{-10pt} 
%	\\	\today 
		\end{minipage} }
\title{
Observation of the \mbox{\boldmath $B_c$} Meson 
in \PbarP\ Collisions at $\sqrt{s} = 1.8$\ TeV
}

\def\r#1{\ignorespaces $^{#1}$}
\author{
%The CDF Collaboration}
% ---------------------- \input{authors} -----------
\font\eightit=cmti8
\hfilneg
\begin{sloppypar}
\noindent
F.~Abe,\r {17} H.~Akimoto,\r {39}
A.~Akopian,\r {31} M.~G.~Albrow,\r 7 A.~Amadon,\r 5 S.~R.~Amendolia,\r {27} 
D.~Amidei,\r {20} J.~Antos,\r {33} S.~Aota,\r {37}
G.~Apollinari,\r {31}
% ...
\end{sloppypar}
\vskip .026in
\begin{center}
(CDF Collaboration)
\end{center}
\vskip .026in
\begin{center}
\r 1  {\eightit Argonne National Laboratory, Argonne, Illinois 60439} \\
\r 2  {\eightit Istituto Nazionale di Fisica Nucleare, University of Bologna,
I-40127 Bologna, Italy} \\
\r 3  {\eightit Brandeis University, Waltham, Massachusetts 02254} \\
\r 4  {\eightit University of California at Los Angeles, Los 
Angeles, California  90024} \\  
\r 5  {\eightit University of Chicago, Chicago, Illinois 60637} \\
\r 6  {\eightit Duke University, Durham, North Carolina  27708} \\
\r 7  {\eightit Fermi National Accelerator Laboratory, Batavia, Illinois 
60510} \\
\r 8  {\eightit University of Florida, Gainesville, FL  32611} \\
\r 9  {\eightit Laboratori Nazionali di Frascati, Istituto Nazionale di Fisica
               Nucleare, I-00044 Frascati, Italy} \\
\r {10} {\eightit University of Geneva, CH-1211 Geneva 4, Switzerland} \\
\r {11} {\eightit Harvard University, Cambridge, Massachusetts 02138} \\
\r {12} {\eightit Hiroshima University, Higashi-Hiroshima 724, Japan} \\
\r {13} {\eightit University of Illinois, Urbana, Illinois 61801} \\
\r {14} {\eightit Institute of Particle Physics, McGill University, Montreal 
H3A 2T8, and University of Toronto,\\ Toronto M5S 1A7, Canada} \\
\r {15} {\eightit The Johns Hopkins University, Baltimore, Maryland 21218} \\
\r {16} {\eightit Institut f\"{u}r Experimentelle Kernphysik, 
Universit\"{a}t Karlsruhe, 76128 Karlsruhe, Germany} \\
\r {17} {\eightit National Laboratory for High Energy Physics (KEK), Tsukuba, 
Ibaraki 305, Japan} \\
\r {18} {\eightit Ernest Orlando Lawrence Berkeley National Laboratory, 
Berkeley, California 94720} \\
\r {19} {\eightit Massachusetts Institute of Technology, Cambridge,
Massachusetts  02139} \\   
\r {20} {\eightit University of Michigan, Ann Arbor, Michigan 48109} \\
\r {21} {\eightit Michigan State University, East Lansing, Michigan  48824} \\
\r {22} {\eightit University of New Mexico, Albuquerque, New Mexico 87131} \\
\r {23} {\eightit The Ohio State University, Columbus, OH 43210} \\
\r {24} {\eightit Osaka City University, Osaka 588, Japan} \\
\r {25} {\eightit Universita di Padova, Istituto Nazionale di Fisica 
          Nucleare, Sezione di Padova, I-35131 Padova, Italy} \\
\r {26} {\eightit University of Pennsylvania, Philadelphia, 
        Pennsylvania 19104} \\   
\r {27} {\eightit Istituto Nazionale di Fisica Nucleare, University and Scuola
               Normale Superiore of Pisa, I-56100 Pisa, Italy} \\
\r {28} {\eightit University of Pittsburgh, Pittsburgh, Pennsylvania 15260} \\
\r {29} {\eightit Purdue University, West Lafayette, Indiana 47907} \\
\r {30} {\eightit University of Rochester, Rochester, New York 14627} \\
\r {31} {\eightit Rockefeller University, New York, New York 10021} \\
\r {32} {\eightit Rutgers University, Piscataway, New Jersey 08855} \\
\r {33} {\eightit Academia Sinica, Taipei, Taiwan 11530, Republic of China} \\
\r {34} {\eightit Texas A\&M University, College Station, Texas 77843} \\
\r {35} {\eightit Texas Tech University, Lubbock, Texas 79409} \\
\r {36} {\eightit Istituto Nazionale di Fisica Nucleare, University of Trieste/
Udine, Italy} \\
\r {37} {\eightit University of Tsukuba, Tsukuba, Ibaraki 315, Japan} \\
\r {38} {\eightit Tufts University, Medford, Massachusetts 02155} \\
\r {39} {\eightit Waseda University, Tokyo 169, Japan} \\
\r {40} {\eightit University of Wisconsin, Madison, Wisconsin 53706} \\
\r {41} {\eightit Yale University, New Haven, Connecticut 06520} \\
\end{center}
}
% ==================================================

\date{\today}
\maketitle
\begin{abstract}
%----------------- \input{abstract} ----------------
We have observed bottom-charm mesons $B_c$\ via 
the decay mode  
$B_c^{\pm} \rightarrow J/\psi \, \ell^{\pm} \nu$\ 
in 1.8~TeV \PbarP\ collisions using the CDF detector 
at the Fermilab Tevatron.
A fit of background and signal contributions 
to the \Jpsil\ mass distribution yielded 
$20.4^{+6.2}_{-5.5}$\ events from $B_c$ mesons.
A fit to   
the same distribution with background alone was rejected 
at the level of 4.8 standard deviations.
We measured 
the $B_c^+$ mass to be $6.40 \pm 0.39 \pm 0.13$ GeV/$c^2$ 
and the $B_c^+$ lifetime to be $\tauBc$.
We measured the production cross section times 
branching ratio for 
\BpJpsilnu\
%$B_c^+ \rightarrow J/\psi \, \ell^+ \nu$\
relative to that for 
$B^+ \rightarrow J/\psi \, K^+$ to be
$\sigmaBr$.
\end{abstract}

%\narrowtext

%----------------  \input{prlmain} -------------------
The \Bcp\ meson is the lowest-mass bound state of a
family of quarkonium states containing a
charm quark and a bottom anti-quark~\cite{prlfoot1}.
Since this pseudoscalar ground state has non-zero flavor, it  
has no strong or electromagnetic decay channels.
It is the last such meson predicted by the Standard Model.
Its weak decay is expected to yield 
a large branching fraction 
to final states containing a 
\Jpsi~\cite{Lusignoli_decay,isgw2,isgw,Chang_decay}.
Non-relativistic potential models 
have been used to predict a \Bc\ mass in the 
range 6.2--6.3 GeV/$c^2$~\cite{Kwong,Eichten}.
In these models, the \cc\ and \bbar\ are tightly bound 
in a very compact system and have a rich spectroscopy of 
excited states.

We expect three major contributions to the \Bc\ decay width:
	$\overline{b} \rightarrow \overline{c} W^+$ 
	with the \cc\ as a spectator, leading to
	final states like $J/\psi \, \pi$\ or $J/\psi \, \ell \nu$;
	$c \rightarrow s W^+$,
	with the \bbar\ as spectator, leading to
	final states like $B_s \, \pi$\ or $B_s \, \ell \nu$;
	and $c \overline{b} \rightarrow W^+$\
	annihilation, leading to final states like 
	$D \, K$, $\tau \, \nu_{\tau}$\ or multiple pions.
Since these processes lead 
to different final states, their amplitudes do not interfere. 
When phase space and other effects are included, 
the predicted lifetime is in the range 0.4--1.4
ps~\cite{Lusignoli_decay,Bigi,Beneke,Gershtein,Colangelo,Quigg}. 
Because of the wide range of predictions, 
a \Bc\ lifetime measurement is a test of the different
assumptions made in the various calculations.
Several authors have also calculated the \Bc\ partial widths
to semileptonic final 
states~\cite{Lusignoli_decay,isgw2,isgw,Chang_decay,Choi}.

The production of \Bc\ mesons has been calculated in perturbative QCD.
At transverse momenta \Pt\ large compared to the \Bc\ mass the
dominant process is that in which a \bbar\ is produced by gluon 
fusion in the hard collision and fragmentation provides the \cc.
At lower \Pt\, a full $\alpha_s^4$\ calculation shows that 
the dominant process is one in which both the \bbar\ and \cc\ 
quarks are produced in the hard scattering.
These calculations~\cite{Lusignoli_prod,Braaten,Chang_prod,Oakes,Masetti}
provide inclusive production cross sections along with distributions in 
\Pt\ and other kinematic variables.

Experimental searches at LEP have obtained limits on
\Bc\ production~\cite{BcDELPHI,BcOPAL,BcALEPH} and a 
few candidate events~\cite{BcOPAL,BcALEPH}.
A prior CDF search placed a limit on \Bc\ production 
in the $B_c^+ \rightarrow J/\psi \, \pi^+$\  decay 
mode~\cite{BcCDF}.

We report here the observation of \Bc\ mesons produced in 
1.8 TeV \pbarp\ collisions at the Fermilab Tevatron collider 
using a 110 \ipb\ data sample collected with the CDF detector.
A more detailed description of this work can
be found in Ref.~\cite{BcPRD}.
We searched for the decay channels \BpJpsimunu\ and \BpJpsienu\ 
followed by \Jpsimumu.
A Monte Carlo calculation of \Bc\ production and 
decay to \Jpsilnu\ showed that,
for an assumed \Bc\ mass of 6.27 GeV/$c^2$, 93\% of 
the \Jpsil\ final state particles would have
\Jpsil\ masses with 
$4.0 < M(J/\psi \, \ell) < 6.0$\ GeV/$c^2$.  
We refer to this as the signal region, 
but we accepted candidates 
with $M(J/\psi \, \ell)$\ between 3.35 and 11 GeV/$c^2$.

We have described the CDF detector in detail elsewhere~\cite{CDF1,CDF2}.
The tracking system for CDF gives a transverse momentum resolution 
$\delta p_T/p_T = [(0.0009\times p_T)^2 + (0.0066)^2]^{1/2}$, 
where \Pt\ is in units of GeV/$c$.
The average track impact parameter resolution 
relative to the beam axis is 
$(13 + (40/p_T))~\mu$m 
in the plane transverse to the beam~\cite{CDF3}.
An online di-muon trigger and subsequent offline 
selection yielded a sample of about 196,000 
$J/\psi \rightarrow \mu^+ \mu^-$ mesons.

The decays we sought, \BpJpsilnu,
have a very simple topology:  
a decay point for $J/\psi \rightarrow \mu^+ \mu^-$\ 
displaced from the primary interaction point
and a third track emerging from the same decay point.  
This \Jpsi\ + track sample included \BpJpsienu, \BpJpsimunu, \BJpsiK, and 
background from various sources.
We subjected the three tracks to a fit that 
constrained the two muons to the \Jpsi\ mass and that
constrained all three tracks to orginate from a common point.
A measure of the time between production and decay 
of a $B_c$ candidate is the quantity
 \ctstar, defined as
\begin{equation} 
 ct^{\ast} = \frac{M(J/\psi \, \ell) \cdot 
  L_{xy}(J/\psi \, \ell)}{|p_T(J/\psi \,\ell)|} 
\label{eq:ctstar}
\end{equation} 
where 
$L_{xy}$ is the distance between the beam centroid and the 
decay point of the
\Bc\ candidate in the  plane perpendicular to the beam
direction  and projected along the direction of the \Jpsil\
combination in that plane,
$M$(\Jpsil) is the mass of the tri-lepton system, and
\Pt(\Jpsil) is its momentum transverse to the beam. 
Our average uncertainty in the measurement of \ctstar\ is 25 \um.
We required $ct^{\ast} >$\ 60 \um. 
%for all candidates in the analysis of the \Bc\ signal significance.

\BJpsiK\ candidates were identified by a  
peak in the $\mu^{+} \mu^{-} K^+$\ mass distribution  
centered at $M(B^+) =  5.279$\ GeV/$c^2$\ with an 
r.m.s. width of 14 MeV/$c^2$. (See Fig. 2 of Ref.~\cite{BcPRD}.)
The peak contained $290 \pm 19$\ events 
after correction for background.
Events within 50 MeV of $M(B^+)$\ were eliminated as 
\BpJpsilnu\ candidates.
 
Electrons were identifed by the association of a 
charged-particle track with $p_T>2$\,GeV$/c$
and an electromagnetic shower in the calorimeter.  
Additional information for identifying electrons
was obtained from specific ionization in the 
tracking chambers and from the shower profile in
proportional chambers embedded in the electromagnetic 
calorimeter.
Muons from \Jpsi\ decay were identifed by matching a 
charged-particle track with $p_T>2$\,GeV$/c$
to a track segment in muon drift
chambers outside the central calorimeter 
(5 to 9 interaction lengths thick depending on angle). 
The third muon was required to have a transverse momentum 
exceeding 3\,GeV/$c$ and to pass through 
an additional three interaction lengths of steel to produce 
a track segment in another set of drift chambers.
We found 23 \BpJpsienu\ candidates of which 19 were in the signal region,
and we found 14 \BpJpsimunu\ candidates of which 12 were in the 
signal region.

Significant contributions to backgrounds 
in the sample of \Bc\ candidates come 
from misidentification of hadron tracks as leptons 
($i.e$. false leptons) and 
from random combinations of real leptons with \Jpsi\ mesons.
There are three significant sources of false leptons:
hadrons that reach the muon detectors without being absorbed;
hadrons that decay in flight into a muon in advance of entering the muon
detectors; and hadrons that are falsely identified as electrons.
In one type of random combination 
external or internal conversions, $i.e.$\  
electrons from photons that convert to $e^+ e^-$\ pairs
in the material around 
the beam line or from Dalitz decay of $\pi^0$.
When the other member of the pair remains undetected,
electrons from these sources contribute 
a ``conversion background'' to the candidate sample.
The other type of random combination involves 
a $B$ that has decayed into a $J/\psi$ 
and an associated $\overline{B}$ that has decayed
semileptonically (or through semileptonic decays of its daughter
hadrons) into a muon or an electron.  
The displaced $J/\psi$ and the lepton can accidentally 
appear to originate from a common point. 
A number of other backgrounds~\cite{BcPRD} were found
to be negligible.
From a combination of data and Monte Carlo calculations,
we determined the \Jpsi\ + track mass 
distribution for each of the sources of background.
As a check of our background calculations, we verified that 
we are able to predict the number of events and mass distribution 
in an independent, background-rich sample of 
same-charge, low-mass lepton pairs.  
(See Fig. 27 in Ref.~\cite{BcPRD}.)

We verified the topology for candidate events by
applying all selection criteria except the requirement that 
the third track intersect the \Jpsi\ vertex.
The impact parameter distribution between the third track
and the \Jpsi\ vertex has a prominent peak at zero,  
demonstrating that, for most candidate events, the three 
tracks arise from a common vertex.  
(See Fig.~28 in Ref.~\cite{BcPRD}.)

\begin{figure}
\epsfxsize=4.4in
\centerline{\epsffile{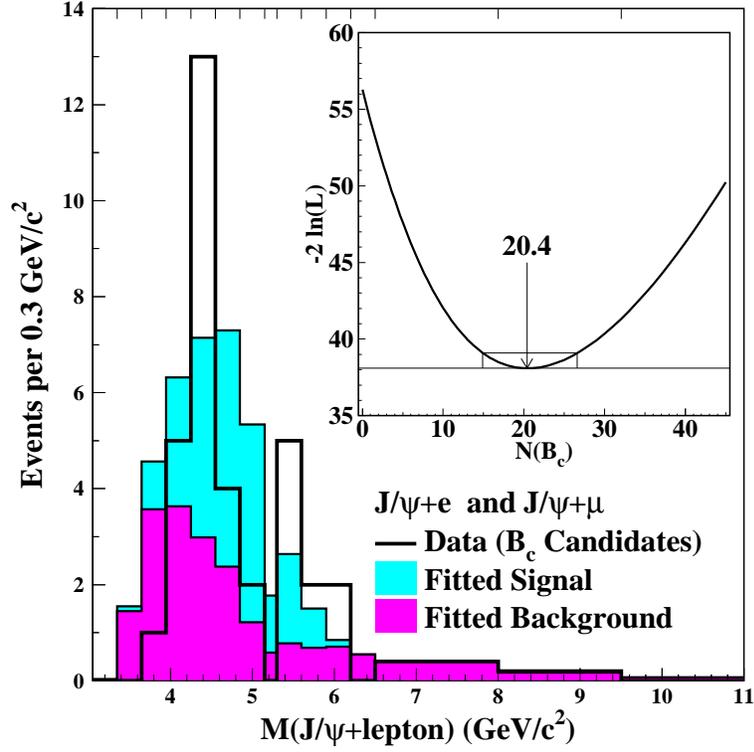}}
\caption{
Histogram of the \Jpsil\  mass that compares 
the signal and background contributions determined in the likelihood fit to 
the combined data for \Jpsie\ and \Jpsimu.
Note that the mass bins vary in width.
The total $B_c$\ contribution is $20.4^{+6.2}_{-5.5}$ events. 
The inset shows the behavior of the log-likelihood 
function $-2 \ln(L)$\ vs. the number of \Bc\ mesons.}
\label{fig:prlf1}
\end{figure}

Table~\ref{tab:like} summarizes the results of the
background calculation and of a simultaneous 
fit for the muon and electron channels 
to the mass spectrum over the region between 
3.35 and 11 GeV/$c^2$~\cite{BcPRD}.
Figure~\ref{fig:prlf1} shows the mass spectra
for the combined \Jpsie\ and \Jpsimu\ candidate 
samples, the combined backgrounds and the fitted 
contribution from \BpJpsilnu\ decay.  The fitted number of 
\Bc\ events is $20.4^{+6.2}_{-5.5}$. 

To test the significance of this result,
we generated a number of Monte Carlo trials with
the statistical properties of the 
backgrounds, but with no contribution from \Bc mesons.
These were subjected to the same fitting procedure
to determine contributions consistent with the
signal distribution arising from background fluctuations.  
The probability of obtaining a yield of 20.4 
events or more is $0.63 \times 10^{-6}$, 
equivalent to a 4.8 standard-deviation effect.  

To check the stability of the \Bc\ signal, 
we varied the value assumed for the  \Bc\ mass.
We generated signal templates, $i.e.$\ 
Monte Carlo samples of \BpJpsilnu,  
with various values of  $M(B_c)$\ from 5.52 to 7.52 GeV/$c^2$.    
The signal template for each value of $M(B_c)$\ and  
the background mass distributions were used 
to fit the mass spectrum for the data.  
This study established that 
the magnitude of the \Bc\ signal is stable over the
range of theoretical predictions for $M(B_c)$, 
and the dependence of the
log-likelihood function on mass yielded 
$M(B_c) = 6.40 \pm 0.39\,{\rm (stat.)} \pm 0.13\,{\rm (syst.)}$\
GeV/$c^2$.
 
\begin{figure}
\epsfxsize=4.4in
\centerline{\epsffile{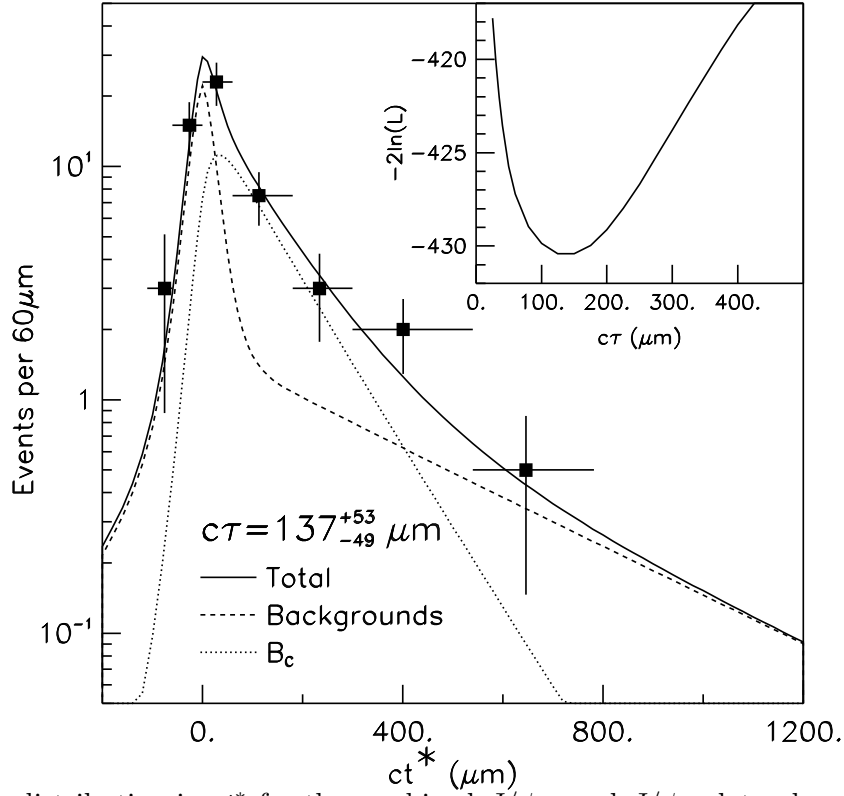}}
\caption{The distribution in \ctstar\  
for the combined \Jpsimu\ and \Jpsie\ data along with 
the fitted curve and contributions to it from signal and background.
The inset shows the log-likelihood function vs. \ctau\ for the \Bc.
}
\label{fig:bc_ct_inset}
\end{figure}

We obtained the mean proper decay length \ctau\ and 
hence the lifetime $\tau$\ of the $B_c$\ meson from 
the distribution of \ctstar.  
We used only events with 
$4.0 < M(J/\psi \, \ell) < 6.0$\ GeV/$c^2$, 
and we changed the threshold requirement on 
\ctstar\ from $c t^* > 60$\ \um\ to $c t^* > -100$\ \um\ 
for this lifetime measurement.  
This yielded a sample of 71 events, 42 \Jpsie\ and 29 \Jpsimu.  
We determined a functional form for the shapes in \ctstar\ for each
of the backgrounds.  
To these, we added a resolution-smeared exponential decay
distribution for a \Bc\ contribution, parametrized by its mean decay
length \ctau.  
Because of the missing neutrino, the proper decay length $ct$\ for 
each event differs from \ctstar\ of Eq.~\ref{eq:ctstar}. 
We convoluted the exponential in $ct$\ with the distribution of 
$ct^{\ast}/ct$\ derived from Monte Carlo studies.
Finally, we incorporated the data from each of the 
candidate events in an unbinned likelihood fit to determine the 
best-fit value of \ctau.  
The data and the signal and background 
distributions are shown in
Fig.~\ref{fig:bc_ct_inset}, and the result is:
\begin{eqnarray}
	c \tau = \ctauBc
\label{eq:ctau}
\end{eqnarray}
\begin{eqnarray}
	  \tau = \tauBc
%\label{eq:tau}
\end{eqnarray}

\begin{figure}
\epsfxsize=4.4in
\centerline{\epsffile{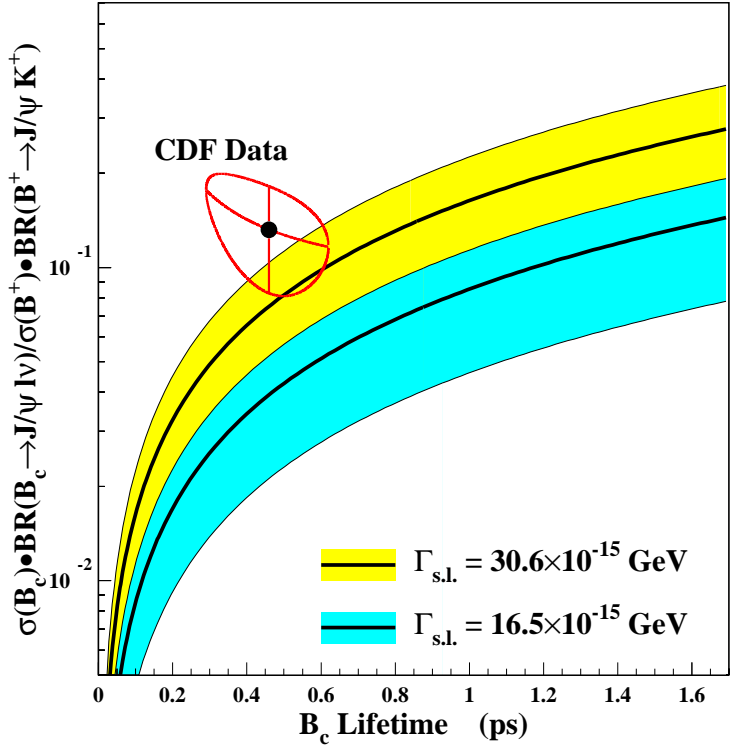}}
\caption{The point with 1-standard-deviation contour 
shows our measured value of the  
$\sigma \cdot BR$\ ratio  
plotted at the value we measure for the \Bc\ lifetime.
The shaded region represents theoretical predictions 
and their uncertainty corridors for two
different values of the semileptonic width 
$\Gamma_{s.l.} = \Gamma(B_c  \rightarrow J/\psi \, \ell \, \nu)$ 
based on Refs.~\protect\cite{Lusignoli_decay} 
and \protect\cite{isgw}.
The other numbers assumed in the theoretical predictions 
are 
$V_{cb} 
	= 0.041 \pm 0.005$~\protect\cite{Particledata},
${\sigma(B_{c}^{+})}/{\sigma(\overline{b})} 
	= 1.3 \times 10^{-3}$~\protect\cite{Lusignoli_prod},
$\frac{\sigma(B^{+})}{\sigma(\overline{b})} 
	= 0.378 \pm 0.022$~\protect\cite{Particledata},
$BR(B^{+} \to J/\psi K^{+}) 
	= (1.01 \pm 0.14) \times 10^{-3}$~\protect\cite{Particledata}.
}
\label{fig:ct_crss1}
\end{figure}

From the 20.4 \Bc\ events and the 290 \BJpsiK\ events, 
we calculated the $B_c$\ 
production cross section times the \BpJpsilnu\ branching fraction
\sigBRBcl\ relative to that for the 
topologically similar decay \BJpsiK. 
Systematic uncertainties 
arising from the luminosity,
from the $J/\psi$\ trigger efficiency,
and from the track-finding efficiencies cancel in the ratio.  
Our Monte Carlo calculations yielded 
the values for the efficiencies that
do not cancel.
The detection efficiency for \BpJpsilnu\ depends on \ctau\ 
because of the requirement that $ct^* > 60$\ \um, and we 
quote a separate systematic uncertainty because of the lifetime
uncertainty.
We assumed that the branching fraction is the same for 
\BpJpsienu\ and \BpJpsimunu. 
We multiply the 20.4 events by a factor $0.85 \pm 0.15$\ 
to correct for other \Bc\ decay channels such as
$B_c \rightarrow \psi^{\prime} \, \ell \nu$~\cite{BcPRD}.
We find
\widetext
\begin{mathletters}
\begin{equation}
	{\cal R}(J/\psi \, \ell \nu) \equiv 
	\frac{\sigma(B_c) \cdot BR(B_c \rightarrow J/\psi \, \ell \nu)}
		   {\sigma(B) \cdot BR(B \rightarrow J/\psi \, K)} 
%\label{eq:calR_def} \\
	= \sigmaBr ,
\label{eq:sigB_ratio}
\end{equation}
\end{mathletters}
\narrowtext
for \Bcp\ and $B^+$\ with 
transverse momenta $p_T > 6.0$\ GeV/$c$\ and rapidities $|y| < 1.0$.
This result is consistent with limits from previous 
searches~\cite{BcDELPHI,BcOPAL,BcALEPH}.
Figure~\ref{fig:ct_crss1} compares phenomenological predictions 
with our measurements of \ctau\ and ${\cal R}(J/\psi \, \ell \nu)$.
Within experimental and theoretical uncertainties, they are consistent.

In conclusion, we report the observation of $B_c$ mesons through
their semileptonic decay modes, $B_c \rightarrow J/\psi \, \ell X$
where $\ell$ is either an electron or a muon.
We measured the \Bc\  mass  and the product 
of its production cross section times 
semileptonic branching fraction which 
confirm phenomenological expectations.
We measured a \Bc\ lifetime consistent with calculations
in which the decay width is dominated by the decay of 
the charm quark.

    We thank the Fermilab staff and the technical staffs of the
participating institutions for their vital contributions.  This work was
supported by the U.S. Department of Energy and National Science Foundation;
the Italian Istituto Nazionale di Fisica Nucleare; the Ministry of Education,
Science and Culture of Japan; the Natural Sciences and Engineering Research
Council of Canada; the National Science Council of the Republic of China; 
the A. P. Sloan Foundation; and the Swiss National Science Foundation.

%\begin{references}
%\bibliography{bc}
%\bibliographystyle{unsrt}
% --------------  insert .bbl file here ------------------

%\end{references}

\newpage

% -------------  \input{prltabdef} -----------------------
\narrowtext
\begin{table}
\caption{$B_c$ Signal and Background Summary}
\label{tab:like} 
\begin{tabular}{ lcc }
 	& \multicolumn{2}{c}{ $3.35 < M(J/\psi \, \ell) <  11.0$\ GeV/$c^2$
	 } \\
\tableline
  & \Jpsie\ Events & \Jpsimu\ Events \\
\tableline
False Electrons	
	& $4.2 \pm 0.4$		& \\
\tableline
Undetected Conversions	& $2.1  \pm 1.7$ 		& \\
\tableline
False Muons
	& & $11.4 \pm 2.4$ \\
\tableline
$B \overline{B}$ bkg. \rule{0pt}{10pt}		
	& $2.3 \pm 0.9$	
	& $1.44 \pm 0.25$ \\
\tableline
Total Background (predicted) 	& $8.6 \pm 2.0$		& $12.8 \pm 2.4$ \\
\hspace{1.2in} (from fit)	& $9.2 \pm 2.0$	&	$10.6 \pm 2.3$ \\
\tableline
Predicted $N$(\BJpsienu)/$N$(\BJpsilnu) 	
	& \multicolumn{2}{c}{$0.58 \pm 0.04$} \\
\tableline
\cline{2-3} 
$e$\ and $\mu$\ Signal (derived from fit)
	& \rule[-5pt]{0pt}{15pt}$12.0^{+3.8}_{-3.2}$	
	& $8.4^{+2.7}_{-2.4}$ \\
Total Signal (fitted parameter)
	& \multicolumn{2}{c}{\rule[-5pt]{0pt}{15pt}$20.4^{+6.2}_{-5.5}$} \\ 
\tableline
  Signal + Background	
\tablenote{ The total number of fitted events was not constrained
to be equal to the number of candidates.}
		& $21.2 \pm 4.3$	& $19.0 \pm 3.5$ \\
Candidates	& 23			& 14 \\
\tableline \tableline
$P$(Null)\rule{0pt}{10pt}
\tablenote{Probability that background alone can fluctuate
	   to produce an apparent signal of 20.4 events or
	   more, based on simulation of statistical fluctuations. }
	& \multicolumn{2}{c}{ $0.63 \times 10^{-6}$ } \\
\end{tabular}
\end{table}

% ------------------ \input{prlfigdef} --------------------
% prl fig 1 prlf1.eps

\newpage
% prl  fig 2 bc_ct_inset.eps

\newpage
% prl  fig 3 ct_crss1.ps

\end{document}